# Energy-band engineering for improved charge retention in fully self-aligned double floating-gate single-electron memories

*Xiaohui Tang†\*, Christophe Krzeminski‡, Aurélien Lecavelier des Etangs-Levallois ‡, Zhenkun Chen‡, Emmanuel Dubois‡, Erich Kasper $, Alim Karmous$, Nicolas Reckinger†, Denis Flandre†, Laurent A. Francis†, Jean-Pierre Colinge\*\* and Jean-Pierre Raskin†*

† Institute of Information and Communication Technologies, Electronics and Applied Mathematics (ICTEAM), Université catholique de Louvain, Place du Levant, 3, B-1348 Louvain-la-Neuve, Belgium.

‡ Institut d'Electronique, de Microélectronique et de Nanotechnologie, IEMN/ISEN UMR CNRS 8520, Avenue Poincaré, Cité Scientifique, F-59652 Villeneuve d'Ascq Cedex, France.

$ University Stuttgart, Inst. Halbleitertechnik, Pfaffenwaldring 47, Stuttgart, D-70569, Germany.

\*\* Tyndall National Institute, University College Cork, Dyke Parade, Cork, Ireland.

\* Institute of Information and Communication Technologies, Electronics and Applied Mathematics (ICTEAM), Université catholique de Louvain, Place du Levant, 3, 1348 Louvain-la-Neuve, Belgium. Tel: 003210477421, Fax: 003210472598, E-mail: xiaohui.tang@uclouvain.be

ABSTRACT. We present a new fully self-aligned single-electron memory with a single pair of nano floating gates, made of different materials (Si and Ge). The energy barrier that prevents stored charge leakage is induced not only by quantum effects but also by the conduction-band offset that arises between Ge and Si. The dimension and position of each floating gate are well defined and controlled. The devices exhibit a long retention time and single-electron injection at room temperature.

KEYWORDS Double floating-gate single-electron memory, heterostructure, quantum effects, conduction-band offset.

To meet strict requirements of low power consumption, fast access speed and high integration density, Si[i], Ge[ii,iii], metallic[iv] and other material[v,vi,vii] nanocrystals (NCs) are considered as the most promising candidates to replace the planar floating gates (FG) in the upcoming generations of flash memory devices. The largest drawback of NC memory devices is the poor charge retention time caused by a very thin tunnel oxide. To fulfill ten-year retention times, double NC floating-gate memory devices were proposed. In the literature, the NCs of the double FG memories are generally made of the same material (either metal[viii,ix] or Si[x,xi]). In the former case, Au and Ag NCs are the most appropriate materials to serve as FGs and Coulomb blockade effects arising between the upper and lower FGs prevent the electron stored in the upper FG from leaking into the channel, therefore improving retention time. However, metallic NCs easily diffuse into the channel region, tunnel and control oxides during the subsequent high-temperature processes and are, therefore, not compatible with CMOS technology. In the latter case, quantum confinement and Coulomb blockade effects (hereafter referred to as "quantum effects") in the lower Si FG induce an energy barrier that prohibits charge leakage from the upper FG to the channel, resulting in a longer retention time. Enabling this quantum effect at room temperature requires the diameter of the lower FG to be less than 5 nm. Furthermore, the two FGs need to have different diameters to avoid having the same energy distribution of the density of states in the two NCs, which would increase tunneling in reverse manner. In order to accurately align the upper and lower NC FGs, Kane[xii] proposed the formation of a lower FG mostly constituted of dopant atoms. Unfortunately, despite of the advances of modern technology, this approach remains unpractical. As far as we know, most of double NC FGs memory devices consist of two layers of NCs, formed by deposition or synthesis. This does not ensure uniform NC diameter and spatial distribution, and totally ignores the alignment of NCs forming a pair of FGs. Although self-aligned double NC FGs obtained by oxidizing the amorphous Si layer below Si NCs[xiii], each resulting device includes many pairs of NCs. The weakness of the multiple NC devices is the NC-to-NC lateral interference and the lateral direct tunneling current, which degrades device performance[xiv]. Particularly, the dispersions of the retention time and memory window due to single electron stochastic behaviors seriously affect the device reliability[xv]. Clearly, the ultimate flash memory device would be a double FG single-electron memory, in which the storage of a single electron represents one bit and each memory cell only involves one set of FGs.

Nowadays, fabricating NCs and nanowires with a width less than 5 nm is not really an issue, but the difficult task is to use them as device components and to place them at the desired location. Here we present a new fabrication method for single-electron memory devices operating at room temperature based on the processing of a $Si/Si_{1-x}Ge_x/Si$ heterostructure (Fig. 1). The device features a single pair of NCs which act as FGs. These two NCs are made of different materials (Si and Ge). Specifically, Si and Ge NCs are selected as the upper FG and the lower FG, respectively. <100> germanium has a smaller electron affinity than <100> silicon[xvi]. This creates a conduction-band offset between the Si and Ge NCs. In addition, the quantum confinement effect is larger in Ge NC because of lower effective mass. These facts translate to a high energy barrier between the upper FG and the channel of the memory device which makes it harder for electrons to leak out of the upper FG. Even if both NCs have the same size or if the lower NC has a diameter larger than 5 nm, the retention time is significantly improved. In our fabrication method, the size of each FG can be precisely defined and controlled by tuning the starting thicknesses of the Si and $Si_{1-x}Ge_x$ layers. It is also important to note that both NCs are automatically self-aligned above the channel of the nanowire transistor in our proposed fabrication process.

The fabrication process is described in Figs. 2a through 2f. Layers of $Si_{0.7}Ge_{0.3}$ and Si layers are successively grown by epitaxy[xvii] on a 50-nm-thick, p-type silicon-on-insulator (SOI) wafer. Both grown layers have a thickness of 20 nm. The resulting heterostructure consists of a $Si/Si_{0.7}Ge_{0.3}/Si$ (20/20/50 nm) stack on a 145-nm-thick buried oxide (BOX), as shown in Fig. 2a. A layer of $Si_3N_4$ with a thickness of 10 nm is deposited on the heterostructure surface to form a hard mask (Fig. 2b). The device layout (Fig. 2c) is composed of a central circle with a diameter of 75 nm connected to wider source and drain regions (200-nm wide) by narrow paths (50-nm wide). The layout is patterned using electron-beam lithography, after which the heterostructure is etched down to the BOX using reactive ion etching (Fig. 2d). Wet oxidation is next carried out at 750°C for 75 min. During this step, both Si NC and Ge NC are simultaneously formed over a Si nanowire (Fig. 2e) in the device centre. The source/drain regions are doped using arsenic implantation at an energy of 140 keV and a dose of $2 \times 10^{14}$ cm$^{-2}$. Dopants are activated by rapid thermal annealing at 950°C for 60 s. Finally, a control gate (Fig. 2f) is deposited and patterned by electron-beam lithography.

Figure 3a shows a scanning electron microscope top view image of a device before the formation of the control gate. Figure 3b is a cross-sectional transmission electron microscopy picture of the device centre, in which a hexagonally shaped Ge NC with a thickness of 20 nm is found. It is vertically self-aligned to both the top Si NC (with a diameter of approximately 5 nm) and the bottom triangular Si nanowire. According to the theory of binary alloy oxidation[xviii], oxide growth from an alloy depends on the alloy composition. For the oxidation of the $Si_{0.7}Ge_{0.3}$ alloy, Si is preferentially oxidized due to the fact that $SiO_2$ has a larger negative Gibbs free energy than $GeO_2$[xix]. Si is thus first oxidized to form $SiO_2$. Ge atoms pile up at the $SiO_2/Si_{0.7}Ge_{0.3}$ interface, forming a Ge-rich shell. Next, Si diffuses through the Ge-rich shell to react with oxygen which has diffused through the $SiO_2$ layer. Ge agglomerates continuously and the thickness of the Ge-rich shell increases during oxidation. On the other hand, the oxygen concentration at the oxidation front decreases with the increase of the $SiO_2$ thickness. As a consequence, Ge itself is not oxidized and condensates instead to form the Ge NC. The growth rate of the control gate oxide (10 nm) and double tunnel oxides is controlled by the density of point defects present at both the $Si_3N_4/Si$ and $Si/Si_{0.7}Ge_{0.3}$ interfaces. These point defects provide reaction sites for the formation of $SiO_2$, thereby increasing the oxidation rate. Double tunnel oxides vertically separate the device centre into a top Si NC, a middle Ge NC and a Si bottom triangular nanowire, which are used as the upper FG, the lower FG and the channel of the present memory device, respectively. The device layout shown in Fig. 2c restricts both FGs to the device centre since the Si and Ge layers are completely consumed by oxidation in the narrow paths (Fig. 3c). The source and drain regions (Fig. 3d), where the 20-nm-thick $Si_{0.7}Ge_{0.3}$ layer appears as a brighter region, are wide enough to avoid the formation of the oxide layers at the interfaces. Thus, the oxidation process only exerts a negligible influence on the wide source and drain regions. As a result, both FGs are self-aligned to the channel and confined to the device center. It is noticeable that the diameter of the central circle in the layout is chosen in such a way that, after dry etching and wet oxidation, the diameter of the Si NC is less than 5 nm to guarantee single-electron injection at room temperature.

As expected, energy dispersive X-ray spectroscopy measurements confirm that the hexagonal lower FG is made of pure Ge as shown in Fig. 4a. The theoretical calculation for the ratio of Si (33.5%) and O atoms (66.5%) also confirms that both tunnel oxides are $SiO_2$ (Fig. 4b). The fabrication process of the self-aligned double FG memory device is simulated by using TSUPREM-4. Figure 4c shows simulation result for the device center. It is clearly seen that the double FG structure is formed during wet oxidation. We also evaluate five main geometrical parameters (two FGs heights, two tunnel oxide thicknesses and control oxide thickness) required to be perfect control in the fabrication of the memory devices. Figure 4d plots the thicknesses of these parameters as a function of oxidation time. It is interesting to note that the height of the lower FG retains the initial thickness of the $Si_{0.7}Ge_{0.3}$ layer (20 nm) during all the oxidation duration. Moreover, after 75 min, the height of the upper FG reaches saturation, self-limiting oxidation[xx] making it difficult to be further reduced. In addition, after 50 min, the thicknesses of the three oxide layers become no sensitive to the oxidation time. Compared to other anisotropic oxidation processes based on dopant enhancement[xxi] or pattern conversion[xxii], both FGs are slowly separated. These facts indicate that the present process is well controlled thanks to the top nitride layer for promoting the lateral oxidation and the $Si_{0.7}Ge_{0.3}$ layer for enabling a tight control of the different geometry parameters. Therefore, our process reproducibility and device reliability are very good so that it can be used for larger area memory fabrication.

In summary, the following distinctive advantages can be identified for the proposed architecture and the associated fabrication process: (i) the energy barrier height for preventing the charge leakage is increased by choosing two different materials for the floating gates, (ii) the size of both nanocrystal floating gates can be precisely controlled and defined by tuning the starting thicknesses of Si and $Si_{0.7}Ge_{0.3}$ layers, (iii) both floating gates and the channel of the nanowire MOSFET are perfectly self-aligned, (iv) the control oxide and the two tunnel oxide layers are simultaneously formed in a single process step, (v) the fabrication process is reproducible and compatible with standard CMOS technology, and (vi) the floating-gate lateral interference effect is fully avoided by using a single pair of nanocrystals. In addition, the proposed concept can be used not only for single/multiple memory device fabrication but also for other applications, such as quantum computing and logic circuits [,xxiii].

An unambiguous signature of the single-electron injection in the FG is the drain current-gate voltage ($I_d$ - $V_g$) hysteretic behavior with quantized threshold voltage shifts, where each current drop/jump corresponds to the injection of a single electron from the channel into the FG or the return of an electron from the FG to the channel[xxiv,xxv]. Figure 5a demonstrates the hysteretic curve of a fully self-aligned double FG single-electron memory at room temperature. When the gate voltage is swept from 0 to 8.5 V, two current drops are observed. They are directly related to electrons from the channel are injected into the FGs. Concerning the charging process of double-layer Si NCs, Lu *et al.*[xxvi] experimentally reported that the electrons were charged by Fowler-Nordheim tunneling so that the electrons were first injected into the lower layer, then into the upper one. However, the simulation results of Yu *et al.*[xxvii] indicated that the electron charge was first realized in the upper dot through the energy level of the lower dot by direct tunneling. In our case, a first electron from the channel is injected into the upper FG through the lower FG at $V_g$ = 7.1 V, giving rise to the first current drop in the $I_d$ - $V_g$ curve. Once this first electron is injected into the lower FG, it is energetically favorable to be injected into the upper one as we will discuss later on. When $V_g$ is further increased to 8.1 V, a second electron is injected into the upper FG by the same way. The condition to observe Coulomb blockade phenomenon in a practical device requires the charging energy ($q^2/2C_\Sigma$) to be at least 3 times larger than the thermal energy ($k_BT$, ~26 meV at 300 K). Ignoring the quantum confinement energy, we estimate the diameter $d$ of a conducting sphere embedded in SiO$_2$ matrix to be 4.7 nm when $q^2/2C_\Sigma = 3k_BT$, where $C_\Sigma$ (= $2\pi\varepsilon_{ox}d$) is the self-capacitance of the sphere[xxviii]. The upper FG diameter of approximately 5 nm in Fig. 3b is similar to the estimated value. This suggests that single-electron charging takes place only in the upper FG since the charging energy corresponding to the lower FG of 20 nm is 18 meV. Even if the quantum confinement energy (13 meV) is taken into account, the characteristic energy (the sum of the quantum confinement energy and the charging energy) of the lower FG is still less than $3k_BT$. When the gate voltage is swept in reverse, only one visible jump is observed in the $I_d$ - $V_g$ curve. This jump is more likely attributed to the return of the electron stored in the lower FG since the positive energy barrier existing between the upper and lower FGs impedes the returning of the electron stored in the upper FG. Yu's simulation results also prove that the retention ability of the lower FG is very limited. It is worth noting that this device is not optimized to reach high performance so that a larger threshold voltage is obtained. Reference devices neither with Si FG nor Ge FG are fabricated by the same process. They do not exhibit any hysteretic phenomena and current drop. This means that the electrons are stored within the FGs or in traps present at the interface of the FGs and $SiO_2$ dielectric[xxix].

To further investigate the single-electron injection mechanism, the time evolution of the drain current ($I_d$ - $t$) is studied. First, a negative pulse of -10 V for 1 s is applied to the control gate to remove the electrons stored in both FGs or set both FGs in their initial state. Then, a constant voltage of 8.5 V is applied to the gate and maintained constant during the measurements. Figure 5b shows one typical $I_d$ - $t$ characteristic curve measured at room temperature. After a short transient, $I_d$ stabilizes at 7.4 nA. When the first electron is injected into the FGs, an abrupt reduction of $I_d$ is observed at $t = 1.5$ s. As time goes by, a second electron is injected into the FGs, at which point a second reduction of $I_d$ appears ($t = 7$ s). The quantization of the current steps, $\Delta I = 0.6$ nA, confirms that the steps are the result of single-electron injection, which can only take place in the upper FG, rather than in the lower one because of its larger size. We also observed the quantization of the current steps in our previous memory device with a Si floating gate of about 5 nm [xxx]. That indicates that the electrons are stored in the upper FG. By repeating the same measurement more than 100 times, we calculated the average injection time of the first electron to be 1 s at 8.5 V, which is of the same order of magnitude as that of our previous memory device, with 0.7 s at 5.58 V.

Figure 6a presents the retention characteristics of the fully self-aligned double FG memory device, measured with ±10 V, 10 s long gate pulses at room temperature. Taking into account the specifications of our device structure (two FGs and a thick control oxide layer), we define the threshold voltage as the gate voltage corresponding to $I_d = 0.5$ nA at $V_d = 500$ mV. During the first ~100 s, there is a degradation of the memory window showing about 50% charge loss (*i.e.* one of the two stored electrons), after which memory window remains stable up to $10^4$ s. It can be speculated that the 50% charge loss of the device originates from the electron stored in the lower FG returning to the close Si channel because the negative energy barrier between the lower Ge FG and Si channel makes it easier for electrons to leak out of the lower FG. To compare the retention time of the double FG memory device with that of the single FG memory device, we also fabricated single FG memory device with a Si FG of about 5 nm and a control oxide of 20 nm. The tunnel oxide thickness of this device is nearly the same as the sum of the two tunnel oxide thickness in the double FG memory device. Figure 6b shows the memory window of the single FG memory device. After $10^4$ s, the memory window width has dropped to 17% of its initial value and tends to zero with increasing time. Concerning the charge loss of the double and single FG memory devices, the former is 50% in $10^4$ s and will keep stable afterwards, while the latter is 83% in $10^4$ s and will tend to zero. This confirms that the lower FG plays an important role in the retention characteristics, which makes charge retention time longer in the single FG.

More importantly, our double FG memory device shows a longer electron retention time compared to the memory device consisting of two layers of Si NCs[13]. The explanation is the following. We consider two double FG single-electron memory devices. For the first one, both FGs are made of Si (Si/Si structure), while for the other one, the upper and lower FGs are made of Si and Ge, respectively (Si/Ge structure). In both structures, the upper FG has the same size of 5 nm as the actual device. All tunnel oxides are supposed to be equally thick. Under these conditions, the electron retention time is determined by the energy barrier height as shown in energy band diagram (Fig. 7a). In the Si/Ge structure, the electron affinity of Ge is smaller than that of Si in the <100> direction . Therefore, the energy barrier height $\Delta E$ for the electron tunneling from the upper FG to the channel is the sum of $\Delta E_Q$ and $\Delta E_B$, where $\Delta E_Q$ is the characteristic energy originating from quantum effects in the lower FG, and $\Delta E_B$ is the conduction-band offset between Ge and Si. In the Si/Si structure, the electrons stored in the upper FG only pass through $\Delta E_Q$ to go back to the channel. But, in the Si/Ge structure (our case), the electrons have to overcome the larger $\Delta E$ barrier. Therefore, the electron tunnel probability is lower, resulting in a longer retention time. To investigate the retention improvement brought by the energy band engineering, we make use of the model reported in ref[13] to calculate a retention improvement factor for both structures. The retention improvement factor $F$ is defined by the ratio of the electron tunnel probability in the single and double FG memory devices and, at a low bias, it is given by:

$$\qquad(1)$$

$$\text{with } \Delta E = \Delta E_B + \Delta E_{Q,lower} - \Delta E_{Q,upper}/2 \qquad(2)$$

where $k_B$ is the Boltzmann constant, $T$ is the temperature (300 K in this work), $\Delta E_B$ is the energy-band offset between two FGs, $\Delta E_{Q,lower}$ and $\Delta E_{Q,upper}$ are the characteristic energies of the lower and upper FGs, respectively. In the Si/Si structure, $\Delta E_B = 0$, while in Si/Ge structure $\Delta E$ is the conduction-band offset. In this calculation, the upper FG size is set to 5 nm as the real device, the Coulomb charging energy is calculated using the electrostatic capacitance of a conducting sphere with a diameter $d$, and the quantum confinement energy is approximated assuming that the nanocrystal is a cube with a size similar to $d$. The calculation results are presented in Fig. 7b. For the Si/Si structure (the line with square markers), we notice that only when the lower FG size is smaller than 5 nm, the retention time is significantly improved compared to the single FG device. But when the lower FG is larger than the upper one, the retention time improvement becomes very limited since $\Delta E$ becomes negative (see Eq. 2). Therefore, the latter case could not serve as a memory cell. Concerning the Si/Ge structure, for a fixed retention improvement factor, a greater conduction-band offset $\Delta E_B$ allows the lower FG size to be larger. Alternately, for a given lower FG size, the greater the conduction-band offset, the greater the retention improvement factor. This argumentation lends strong support to our experimental results and greatly relaxes the size constraints on the lower floating gate.

In conclusion, we report a new method for fabricating fully self-aligned double floating-gate single-electron memory devices using a Si/Si$_{1-x}$Ge$_x$/Si multilayered structure. The fabricated device contains only a single pair of floating gates: an upper Si nanocrystal floating gate (~5 nm) and a lower Ge nanocrystal floating gate (~20 nm). Both floating gates are automatically self-aligned to the channel of the nanowire transistor. The size of each floating gate can be precisely controlled by modulating the thickness of the initial Si and Si$_{1-x}$Ge$_x$ layers and self-limiting oxidation. The energy barrier that prevents charge leakage from the nanocrystals arises both from quantum effects and from the conduction-band offset that exists between Ge and Si. This offset significantly increases the retention time of the memory device. This method opens a new perspective for future designs of multiple-level memory cells and quantum computers. The proposed memory device demonstrates two decisive advantages: a long retention time and a single-electron injection at room temperature.

ACKNOWLEDGMENT. We thank all the engineers and technicians in the UCL-WINFAB clean room facilities for their technical support. This work was supported by the NANOSIL European Network of Excellence under Contract 216171. Xiaohui Tang is a senior researcher of the F.R.S.-F.N.R.S.

**Figure 1.** Schematic view of a fully self-aligned double floating-gate single-electron memory. The upper floating gate is a Si nanocrystal and the low floating gate is a Ge nanocrystal. Both floating gates are fully self-aligned to the channel of the nanowire MOSFET.

**Figure 2.** Processing steps for fabricating the fully self-aligned double floating-gate single-electron memory. **a**, A heterostructure composed of a Si/Si$_{0.7}$Ge$_{0.3}$/Si on a buried oxide (BOX), obtained by epitaxy. **b**, A layer of Si$_3$N$_4$ is deposited on the heterostructure surface to form a hard mask. **c**, The device layout is composed of a central circle connected to wider source and drain regions by narrow paths. **d**, The heterostructure is etched down to the BOX by reactive ion etching. **e**, The upper Si floating

gate and lower Ge floating gate are formed over the channel of the nanowire MOSFET by wet oxidation.

**f**, The control gate electrode is formed by a polysilicon or a metallic layer.

**Figure 3.** Electron microscope pictures of the device. **a**, Scanning electron microscope (SEM) top view of the memory device before the formation of the control gate. The central circle is connected to wider source and drain regions by narrow paths. **b**, Transmission electron microscope (TEM) cross-sectional picture of the center of the device showing the upper Si floating gate, the lower Ge floating gate and the triangular Si channel of nanowire MOSFET. **c**, SEM cross-sectional picture of the path region of the memory device, constituted of the Si triangular nanowire only. **d**, SEM cross-sectional picture of the memory device in the source and drain regions. These regions are made of a Si/ $Si_{0.7}Ge_{0.3}$ /Si stack.

**Figure 4.** Energy dispersive X-ray spectroscopy (EDS) spectra and process simulation results. **a**, EDS for the lower floating gate with hexagonal shape, showing that it is made of pure Ge atom. **b**, EDS for both tunnel oxides, indicating that it is $SiO_2$ layer. **c.** Simulated device center based on TSUPREM-4. **d**. Five main geometrical parameter evolutions of the double floating gate memory device (two FG heights, two tunnel oxide thicknesses and control oxide thickness) as a function of oxidation time at 750°C.

**Figure 5.** Electrical characteristics of a fully self-aligned double floating-gate single-electron memory device at room temperature. **a**, Hysteretic behavior at $V_d$ = 500 mV, showing two/one drain current drops/jump when the gate voltage is swept forth/back. **b**, Time evolution of drain current at $V_g$ = 8.5 V and $V_d$ = 500 mV, showing the quantized current steps.

**Figure 6.** Retention properties of the devices with single Si floating gate and double floating gates (Si/Ge) at room temperature. **a**, Memory window of the double floating-gate memory device with programming/erasing at ±10 V for 10 s. The threshold voltage was defined at $I_d$ = 0.5 nA and $V_d$ = 500

mV. After 100 s, the memory window width gets mostly saturated with time. **b**, Memory window of the single floating-gate memory device with programming/erasing at ±10 V for 10 s. After $10^4$ sec, the memory window width has dropped 17% of its initial value and tends to zero with increasing time.

**Figure 7.** Energy-band diagram of the double floating-gate memory and the retention improvement factor for the double floating-gate devices with either two Si floating gates (Si/Si) or the combination of Si and Ge floating gates (Si/Ge). **a**, Energy band diagram in a low or zero gate voltage configuration. The energy barrier $\Delta E$ is the sum of the conduction-band offset $\Delta E_B$ between Ge and Si and the characteristic energy $\Delta E_Q$ induced by Coulomb blockade and quantum confinement effects. **b**, Calculated retention improvement factors for two types of structures (Si/Si and Si/Ge) as a function of the lower floating gate dimension $d$.

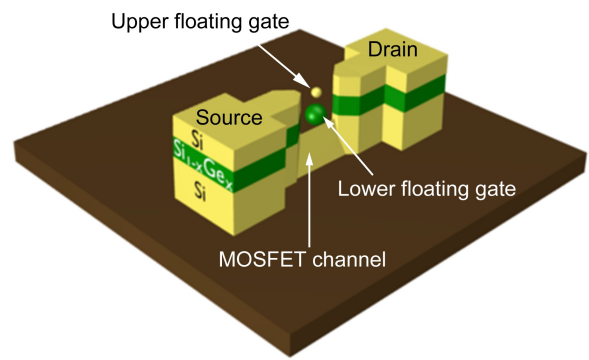

Fig. 1

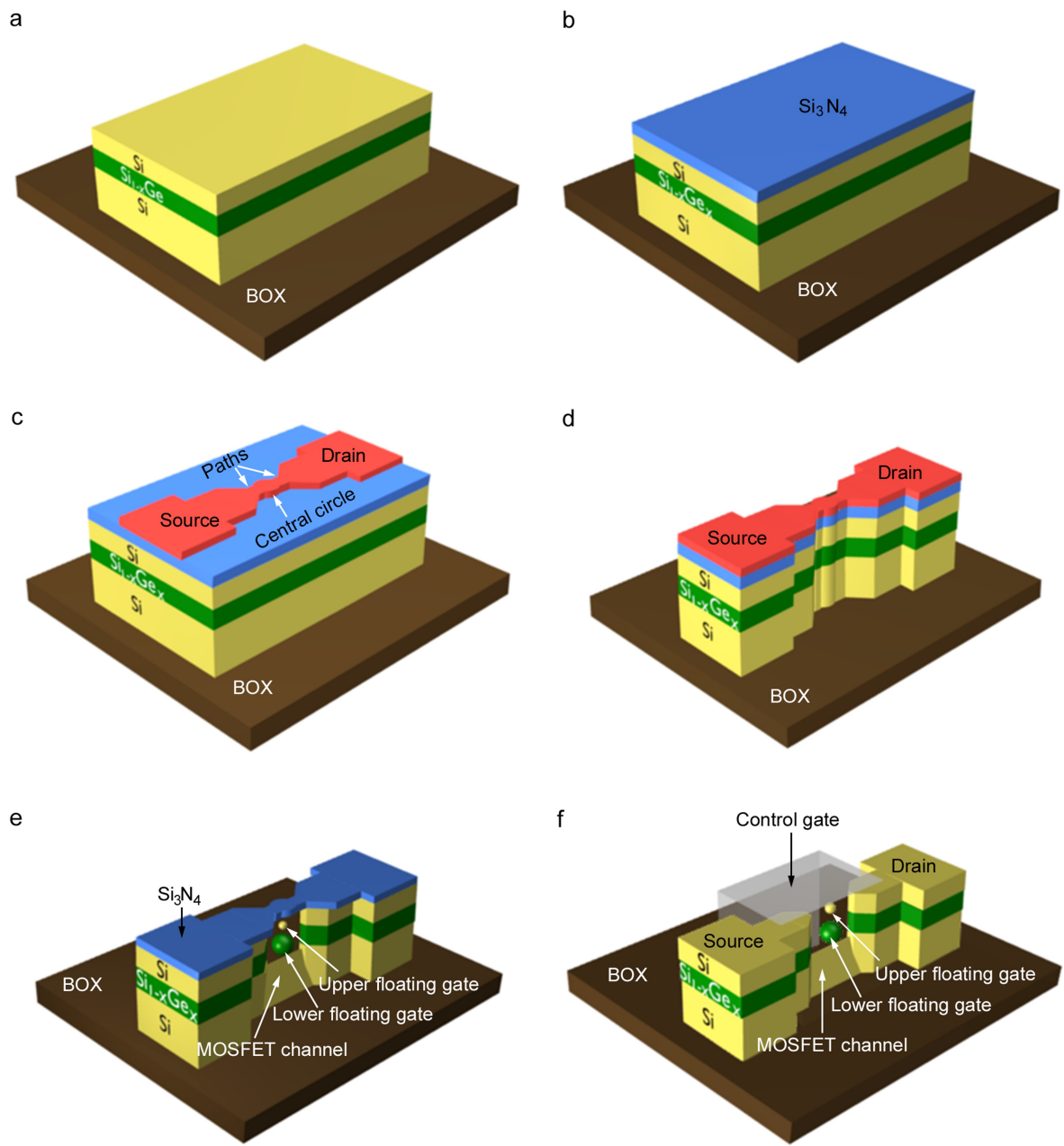

Fig. 2

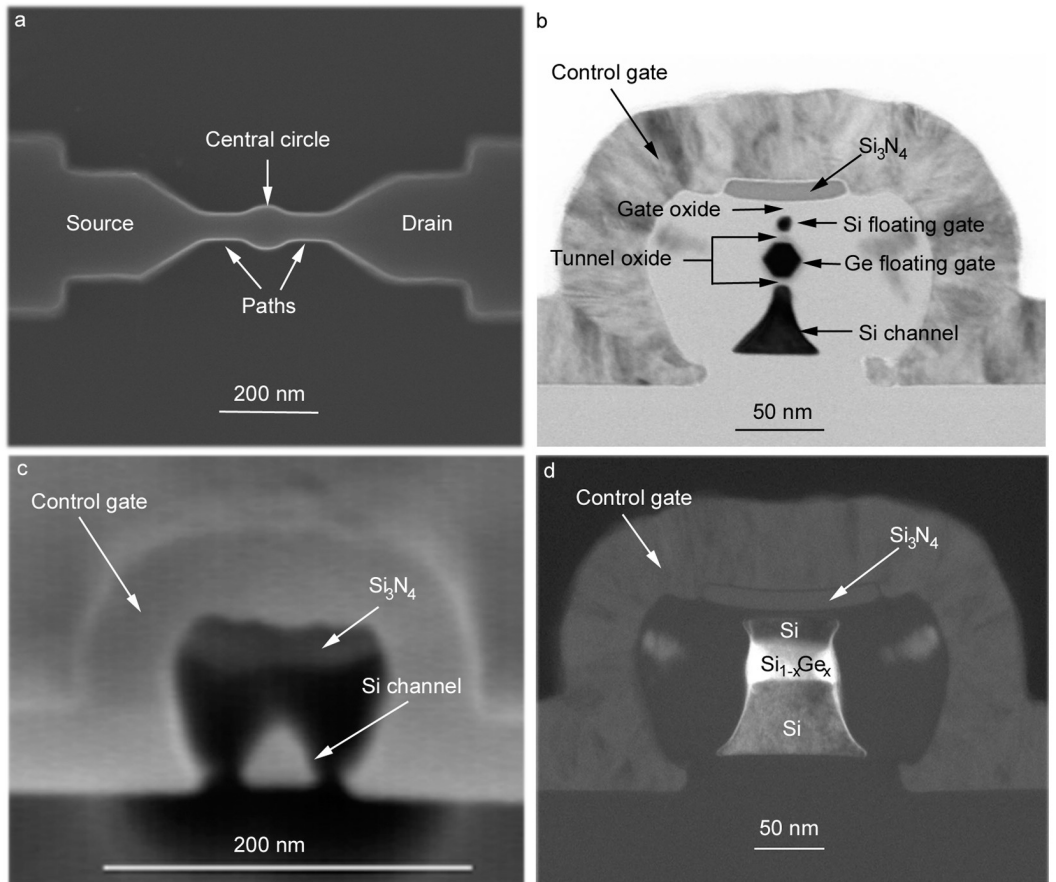

Fig. 3

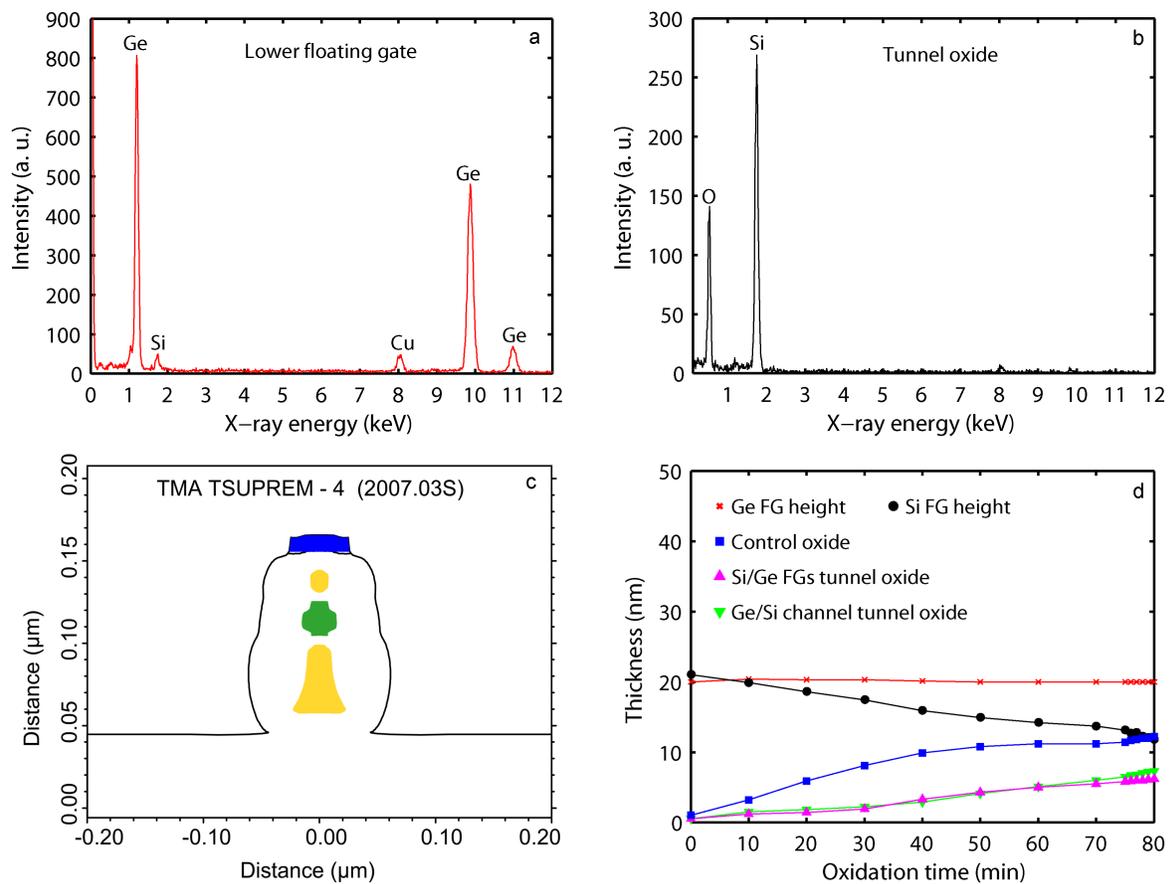

Fig. 4

Fig. 5

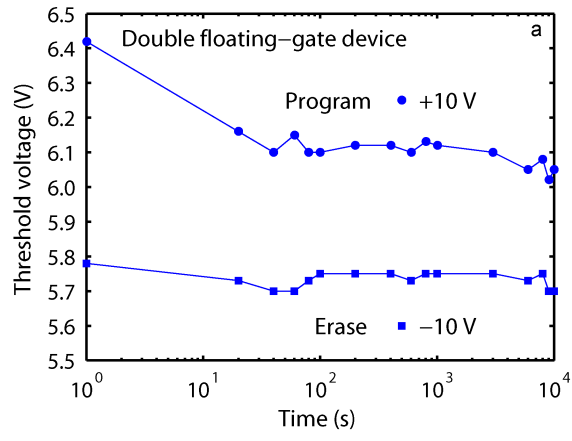 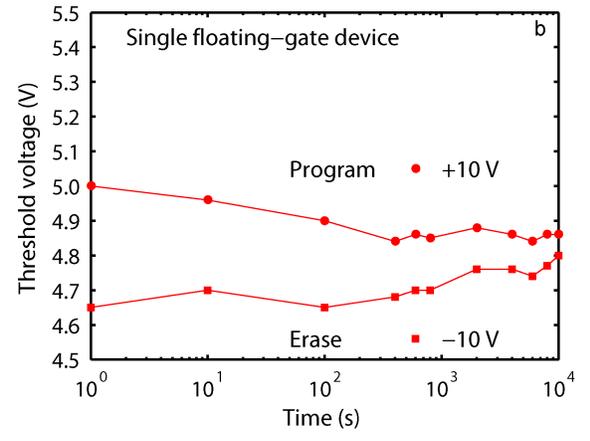

Fig. 6

Fig. 7